\documentclass[12pt]{article}

\usepackage{a4wide}
\usepackage{hyperref}
\usepackage{graphicx}
\usepackage{amsmath}
\usepackage{amssymb}
\usepackage{floatpag}  

\newcommand{\GeV}{\,\text{GeV}}
\newcommand{\TeV}{\,\text{TeV}}

\newcommand{\PU}{\text{PU}}

\newcommand{\cut}{\text{cut}}
\newcommand{\ptcut}{p_t^\cut}
\newcommand{\nPU}{n_\text{PU}}

\newcommand{\order}[1]{{\cal O}\left(#1\right)}

\usepackage{color}

\title{SoftKiller, a particle-level pileup removal method}
\author{Matteo Cacciari,$^{1,2,3}$ Gavin P. Salam$^{4,}$\footnote{On
    leave from CNRS, UMR 7589, LPTHE, F-75005, Paris, France.}\;  and Gregory
  Soyez$^{5}$\\\footnotesize
\footnotesize$^1$Universit\'e Paris Diderot, Paris, France\\
\footnotesize$^2$Sorbonne Universit\'es, UPMC Univ Paris 06, UMR 7589, LPTHE, F-75005, 
Paris, France\\
\footnotesize$^3$CNRS, UMR 7589, LPTHE, F-75005, Paris, France\\
\footnotesize$^4$CERN, PH-TH, CH-1211 Geneva 23, Switzerland\\
\footnotesize\mbox{$^5$IPhT, CEA Saclay, CNRS URA 2306, F-91191
  Gif-sur-Yvette Cedex, France}
}

\begin{document}
\maketitle
\vspace{-10cm}
\begin{flushright}
  CERN-PH-TH/2014-116\\
  July 2014
\end{flushright}
\vspace{8cm}

\begin{abstract}
  Existing widely-used pileup removal approaches correct the momenta
  of individual jets.
  In this article we introduce an event-level, particle-based pileup
  correction procedure, SoftKiller.
  It removes the softest particles in an event, up to a transverse
  momentum threshold that is determined dynamically on an
  event-by-event basis.
  In simulations, this simple procedure appears to be reasonably
  robust and brings superior jet resolution performance compared to
  existing jet-based approaches.
  It is also nearly two orders of magnitude faster than methods based
  on jet areas.
\end{abstract}

\section{Introduction}

At high-luminosity hadron colliders such as CERN's Large Hadron
Collider (LHC), an issue that has an impact on many analyses is pileup, the
superposition of multiple proton-proton collisions at each bunch
crossing.
Pileup affects a range of observables, such as jet momenta and shapes,
missing transverse energy and lepton and photon isolation.
In the specific case of jets, it can add tens of GeV to a jet's
transverse momentum and significantly worsens the resolution for
reconstructing the jet momentum.
In the coming years the LHC will move towards higher luminosity
running, ultimately increasing pileup by up to a factor of ten for the
high-luminosity LHC~\cite{HLLHC}.
The experiments' ability to mitigate pileup's adverse effects will
therefore become increasingly crucial to fully exploit the LHC
data, especially at low and moderate momentum scales, for example in
studies of the Higgs sector.

Some approaches to reducing the impact of pileup are deeply rooted in
experimental reconstruction procedures.
For example, charged hadron subtraction (CHS) in the context of
particle flow~\cite{pflow}, exploits detectors' ability to identify
whether a given charged track is from a pileup vertex or not.
Other aspects of pileup mitigation are largely independent of the
experimental details: for example both ATLAS and
CMS~\cite{ATLAS-PU-Performance,CMS-PU-Performance} rely on the
area--median approach~\cite{areasub,areas}, which makes a global
estimate for the transverse momentum-flow density, $\rho$, and then
applies a correction to each jet in proportion to its area.

In this article, we introduce and study a new generic pileup-removal
method. 
Instead of correcting individual jets, it corrects for pileup at the
level of particles.
Such a method should make a guess, for each particle in an event, as
to whether it comes from pileup or from the hard collision of
interest.
Particles deemed to be from pileup are simply discarded, while the
much smaller set of residual ``hard-collision'' particles are passed
to the jet clustering.
Event-wide particle-level subtraction, if effective, would greatly
simplify pileup mitigation in advanced jet studies such as those that
rely on jet substructure~\cite{boost2012}.
Even more importantly, as we shall see, it has the potential to bring
significant improvements in jet resolution and computational speed.
This latter characteristic makes our approach particularly appealing
also for trigger-level applications.

The basis of our pileup suppression method, which we dub
``SoftKiller'' (SK), is that the simplest characteristic of a particle that
affects whether it is likely to be from pileup or not is its transverse
momentum.
In other words, we will discard particles that fall below a certain transverse
momentum threshold.
The key feature of the method will be its event-by-event determination
of that threshold,
chosen as the lowest $p_t$ value
that causes $\rho$, in the median--area method, to be evaluated as
zero.
In a sense, this can be seen as the extreme limit of ATLAS's approach
of increasing the topoclustering noise threshold as pileup
increases~\cite{Loch@PUWS2014}.

This approach might at first sight seem excessively na\"ive in
its simplicity.
We have also examined a range of other methods.
For example, one approach involved an all-orders matrix-element
analysis of events, similar in spirit to shower
deconstruction~\cite{Soper:2011cr}; others involved event-wide extensions of
a recent intrajet particle-level subtraction method~\cite{Berta:2014eza}
and subjet-level~\cite{quality,Krohn:2013lba} approaches (see also
\cite{Cacciari:2014jta}); we have also been inspired by
calorimeter~\cite{Kodolova} and particle-level~\cite{YueShi} methods
developed for heavy-ion collisions.
Such methods and their extensions have significant potential.
However we repeatedly encountered additional complexity, for example
in the form of multiple free parameters that needed fixing, without a
corresponding gain in performance.
Perhaps with further work those drawbacks can be alleviated, or
performance can be improved.
For now, we believe that it is useful to document one method
that we have found to be both simple and effective.

\section{The SoftKiller method}
\label{sec:area-median-recall}

The SoftKiller method involves eliminating particles below some $p_t$
cutoff, $\ptcut$, chosen to be the minimal value that ensures that $\rho$ is
zero.
Here, $\rho$ is the event-wide estimate of transverse-momentum flow
density in the area--median approach~\cite{areasub,areas}: the event
is broken into patches and $\rho$ is taken as the median, across all
patches, of the transverse-momentum flow density per unit area in
rapidity-azimuth:
\begin{equation}
  \label{eq:rho-median-no-rap}
  \rho = \underset{i \in \text{patches}}{\text{median}} \left\{ \frac{p_{ti}}{A_i}\right\}\,,
\end{equation}
where $p_{ti}$ and $A_{i}$ are respectively the transverse momentum
and area of patch $i$.
In the original formulation of the area--median method, the patches
were those obtained by 
running inclusive $k_t$ clustering~\cite{kt}, but subsequently it was
realised that it was much faster and equally effective to use (almost)
square patches of size $a\times a$ in the rapidity-azimuth plane.
That will be our choice here.
The use of the median ensures that hard jets do not overly bias
the $\rho$ estimate (as quantified in~Ref.~\cite{Cacciari:2009dp}).%
\footnote{One practically important aspect of the area--median method
  is the significant rapidity dependence of the pileup, most easily
  accounted for through a manually determined rapidity-dependent
  rescaling. 
  This is discussed in detail in appendix~\ref{sec:rap-dep}.}

\begin{figure}
  \centering
  \includegraphics[width=0.48\textwidth]{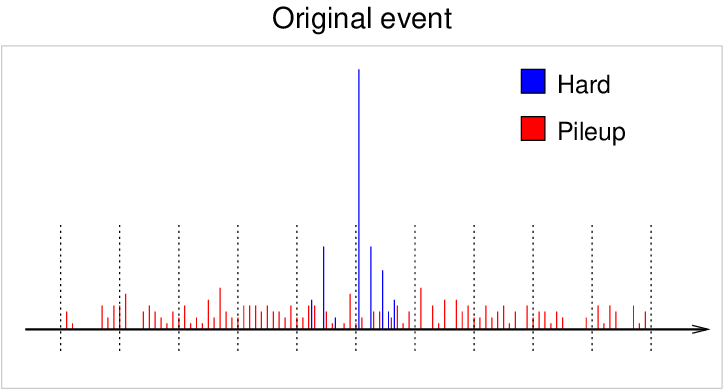}%
  \hfill
  \includegraphics[width=0.48\textwidth]{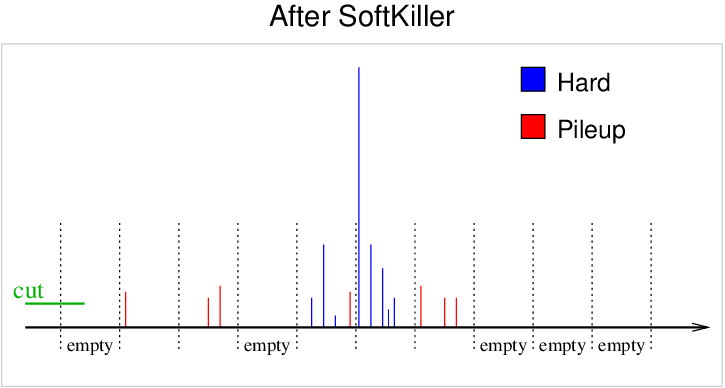}%
  \caption{Illustration of the SoftKiller method. The left plot
    depicts particles in an event, with the hard event particles shown in blue
    and the pileup particles shown in red. On the right, the same
    event after applying the SoftKiller. The vertical dotted lines
    represent the edges of the patches used to estimate the pileup
    density $\rho$.} 
  \label{fig:SK-illustration}
\end{figure}

Choosing the minimal transverse momentum threshold, $p_t^\cut$,
that results in $\rho = 0$ is equivalent to gradually raising the
$p_t$ threshold until exactly half of the patches contain no
particles, which ensures that the median is zero.
This is illustrated in Fig.~\ref{fig:SK-illustration}.
Computationally, $p_t^\cut$ is straightforward to evaluate: one
determines, for each patch $i$, the $p_t$ of the hardest particle in
that patch, $p_{ti}^{\max}$ and then $p_t^\text{cut}$ is given by the 
median of $p_{ti}^{\max}$ values:
\begin{equation}
  \label{eq:ptmin}
  p_t^\cut = \underset{i \in \text{patches}}{\text{median}} \left\{ p_{ti}^{\max} \right\}\,.
\end{equation}
With this choice, half the patches will contain only particles that
have $p_t < \ptcut$. 
These patches will be empty after application of the $p_t$ threshold,
leading to a zero result for $\rho$ as defined in
Eq.~(\ref{eq:rho-median-no-rap}).\footnote{Applying a $p_t$ threshold
  to individual particles is not collinear safe; in the specific
  context of pileup removal, we believe that this is not a significant
  issue, as we discuss in more detail in
  Appendix~\ref{sec:collinear-safety}. }
The computational time to evaluate $\ptcut$ as in Eq.~(\ref{eq:ptmin})
scales linearly in the number of particles and the method should be
amenable to parallel implementation.

Imposing a cut on particles' transverse momenta eliminates most of the
pileup particles, and so might reduce the fluctuations in residual pileup
contamination from one point to the next within the event.
However, as with other event-wide noise-reducing pileup and underlying-event
mitigation approaches, notably the CMS heavy-ion
method~\cite{Kodolova} (cf.\ the analysis in Appendix A.4 of
Ref.~\cite{Cacciari:2011tm}), the price that one pays for noise
reduction is the introduction of biases.
Specifically, some particles from pileup will be above
$p_t^\cut$ and so remain to contaminate the jets, inducing a
net positive bias in the jet momenta.
Furthermore some particles in genuine hard jets will be lost, because
they are below the $p_t^\cut$, inducing a negative bias in the
jet momenta.
The jet energy scale will only be correctly reproduced if these two
kinds of bias are of similar size,\footnote{For patch areas that are
  similar to the typical jet area, this can be expected to happen
  because half the patches will contain residual pileup of order
  $\ptcut$, and since jets tend to have only a few low-$p_t$
  particles from the hard scatter, the loss will also be order of
  $\ptcut$.} so that they largely cancel.
There will be an improvement in the jet resolution if the fluctuations in
these biases are modest.

\begin{figure}
  \includegraphics[width=0.48\textwidth]{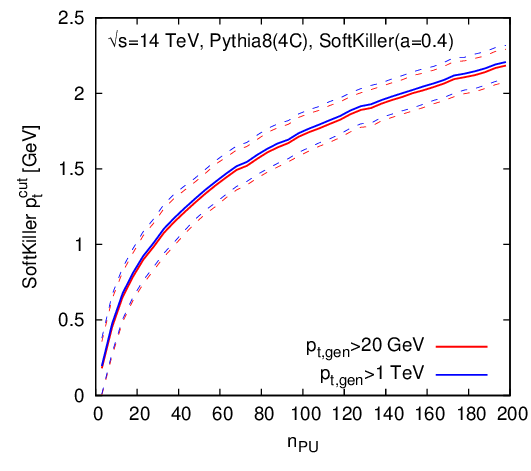}
  \hfill
  \includegraphics[width=0.48\textwidth]{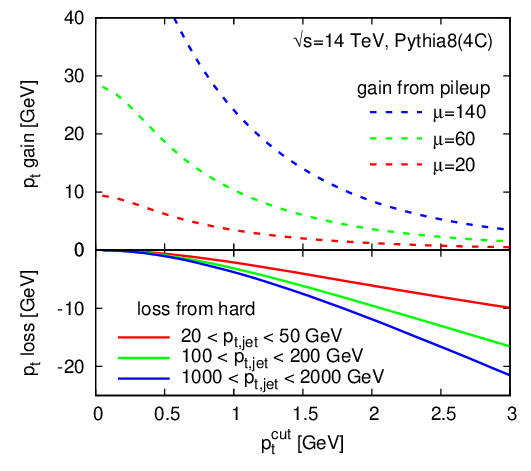}
  \caption{Left: value of the $p_t$ cut applied by the SoftKiller,
    displayed as a function of the number of pileup events. We show
    results for two
    different values of the generator minimal $p_t$ for the hard
    event, $p_{t,\rm gen}$. The solid line is the average $\ptcut$ value while
    the dashed lines indicate the one-standard-deviation band.
    Right: plot of the $p_t$ that is lost when applying a given
    $p_t$ cut (the $x$ axis) to the constituents of jets
    clustered (anti-$k_t$, $R=0.4$) from the hard event (solid lines)
    and the residual pileup $p_t$ that remains after applying that
    same cut to the constituents of circular patches of radius $0.4$ in
    pure-pileup events (dashed lines).}
  \label{fig:killer-cut}
\end{figure}

Figure~\ref{fig:killer-cut} shows, on the left, the average
$p_t^\cut$ value, together with its standard deviation (dashed
lines), as a function of the number of pileup interactions, $n_\PU$.
The event sample consists of a superposition of $n_\PU$ zero-bias on
one hard dijet event, in 14~TeV proton--proton collisions, all
simulated with Pythia~8 (tune 4C)~\cite{Pythia8}.
The 4C tune gives reasonable agreement with a wide range of
minimum-bias data, as can be seen by consulting
MCPlots~\cite{Karneyeu:2013aha}.\footnote{In appendix
  \ref{sec:tune-dep} we also briefly
  examine the Pythia~6~\cite{Sjostrand:2006za} Z2
  tune~\cite{Field:2011iq}, and find very similar results.}
The underlying event in the hard event has been switched off, and all
particles have been made massless, maintaining their $p_t$, rapidity
and azimuth.\footnote{If one keeps the underlying event in the hard
  event, much of it (about $1\GeV$ for both the area--median approach
  and the SoftKiller) is subtracted together with the pileup
  correction, affecting slightly the observed shifts. Keeping massive
  particles does not affect the SK performance but requires an extra
  correction for the area--median subtraction~\cite{Soyez:2012hv}. We
  therefore use massless particles for simplicity.}
These are our default choices throughout this paper.
The grid used to determine $p_t^\cut$ has a spacing of $a \simeq 0.4$
and extends up to $|y| < 5$. 
One sees that $p_t^\cut$ remains moderate, below $2\GeV$, even
for pileup at the level foreseen for the high-luminosity upgrade of
the LHC (HL-LHC), which is expected to reach an average
(Poisson-distributed) number of 
pileup interactions of $\mu \simeq 140$.
The right-hand plot shows the two sources of bias: the lower (solid)
curves, illustrate the bias on the hard jets induced by the loss of
genuine hard-event particles below $p_t^\cut$.
Jet clustering is performed with the anti-$k_t$ jet
algorithm~\cite{Cacciari:2008gp} with $R=0.4$, as implemented in a
development version of
FastJet~3.1~\cite{Cacciari:2005hq,FastJet}.\footnote{For our purposes
  here, the version that we used is equivalent to the most recent
  public release, FastJet~3.0.6.}
The three line colours correspond to different jet $p_t$ ranges. 
The loss has some dependence on the jet $p_t$ itself, notably for
higher values of $p_t^\cut$.%
\footnote{In a local parton-hadron duality type approach to calculate
  hadron spectra, the spectrum of very low $p_t$ particles in a jet of
  a given flavour
  is actually independent of the jet's $p_t$~\cite{KLO}.}
In particular it grows in absolute terms for larger jet $p_t$'s,
though it decreases relative to the jet $p_t$.
The positive bias from residual pileup particles (in circular patches
of radius $0.4$ at rapidity $y=0$) is shown as dashed curves, for
three different pileup levels.
To estimate the net bias, one should choose a value for $n_\PU$,
read the average $p_t^\cut$ from the left-hand plot, and for that
$p_t^\cut$ compare the solid curve with the dashed curve that
corresponds to the given $n_\PU$.
Performing this exercise
reveals that there is
indeed a reasonable degree of cancellation between the positive and
negative biases.
Based on this observation, we can move forward with a more detailed
study of the performance of the method.\footnote{A study of fixed
  $p_t$ cutoffs, rather than dynamically determined ones, is performed
  in Appendix~\ref{fixed-pt-cutoff}.}

\section{SoftKiller performance}
\label{sec:performance}

For a detailed study of the SoftKiller method, the first step is to
choose the grid spacing $a$ so as to break the event into patches.
The spacing $a$ is the one main free parameter of the method. 
The patch-size parameter\footnote{or $k_t$ jet radius.} is present also
for area--median pileup subtraction.
There the exact choice of this parameter is not too critical.
The reason is that the median is quite stable when pileup levels are
high: all grid cells are filled, and nearly all are dominated by
pileup.
However the SoftKiller method chooses the $p_t^\cut$ so as
to obtain a nearly empty event. 
In this limit, the median operation becomes somewhat more sensitive to
the grid spacing~\cite{Cacciari:2009dp}.

\begin{figure}
  \thisfloatpagestyle{empty}
  \includegraphics[width=\textwidth]{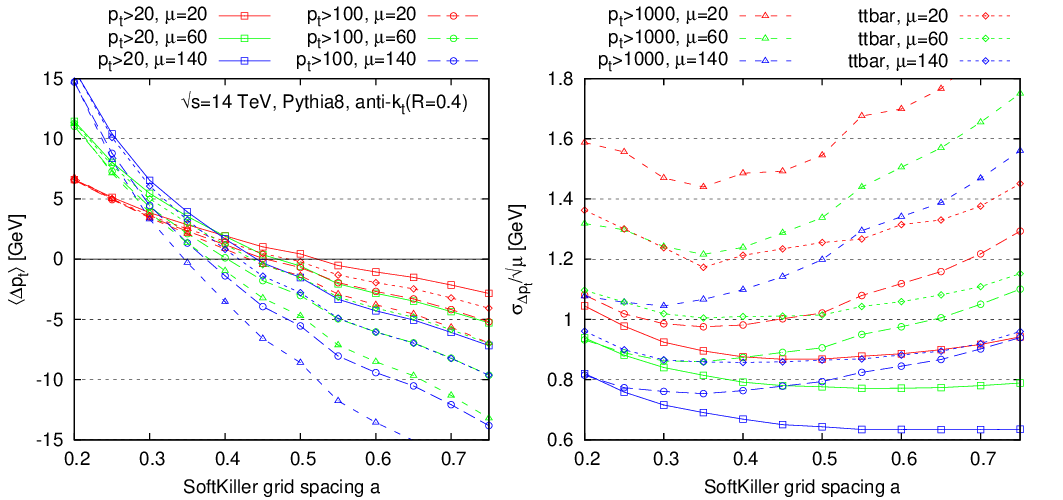}
  \caption{Scan of the SoftKiller performances as a function the
    grid-spacing parameter $a$ for different hard-event samples and 
    three different pileup levels (Poisson-distributed with average
    pileup event multiplicities of $\mu=20,60,140$).
    Left: average $p_t$ shift; right: $p_t$ shift dispersion, normalised
    to $\sqrt{\mu}$ for better readability.
    %
    Results are given for a variety of hard processes
    and pileup conditions to illustrate robustness. 
    Curves labelled $p_t > X$ correspond to dijet events, in which one
    studies only those jets that in the hard event have a transverse
    momentum greater than $X$.
    For the $t\bar t$ sample, restricted to fully hadronic decays of
    the top quarks, the study is limited to jets that have $p_t > 50
    \GeV$ in the hard event.  }
  \label{fig:killer-scan}
\end{figure}

Fig.~\ref{fig:killer-scan} considers a range of hard event samples
(different line styles) and pileup levels (different colours). 
For each, as a function of the grid spacing $a$, the left-hand plot
shows the average, $\langle \Delta p_t\rangle$, of the net shift in
the jet transverse momentum,
\begin{equation}
  \label{eq:Delta-pt}
  \Delta p_t = p_t^\text{corrected} -
  p_t^\text{hard} \,,
\end{equation}
while the right-hand plot shows the dispersion, $\sigma_{\Delta p_t}$,
of that shift from one jet to the next, here normalised to
$\sqrt{\mu}$ (right).

One sees that the average jet $p_t$ shift has significant dependence on
the grid spacing $a$. 
However, there exists a grid spacing, in this case $a\simeq 0.4$,
for which the shift is not too far from zero and not too dependent
either on the hard sample choice or on the level of pileup. 
In most cases the absolute value of the shift is within about $2\GeV$,
the only exception being for the $p_t > 1000\GeV$ dijet sample, for which
the bias can reach up to $4\GeV$ for $\mu=140$.
This shift is, however, still less than the typical best experimental
systematic error on the jet energy scale, today of the order of
$1\%$ or slightly better~\cite{Kirschenmann:2012toa,Aad:2014bia}.

It is not trivial that there should be a single grid spacing that is
effective across all samples and pileup levels: the fact that there is
can be considered phenomenologically fortuitous.
The value of the grid spacing $a$ that minimises the typical shifts is
also close to the value that minimises the dispersion in the
shifts.\footnote{In a context where the net shift is the sum of two
  opposite-sign sources of bias, this is perhaps not too surprising:
  the two contributions to the dispersion are each likely to be of the
  same order of magnitude as the individual biases, and their sum
  probably minimised when neither bias is too large.}
That optimal value of $a$ isn't identical across event samples, and
can also depend on the level of pileup.
However the dispersion at $a = 0.4$ is always close to the actual
minimal attainable dispersion for a given sample.
Accordingly, for most of the rest of this article, we will work with a grid
spacing of $a=0.4$.\footnote{A single value of $a$ is adequate as long
  as jet finding is carried out mostly with jets of radius $R \simeq
  0.4$. Later in this section we will supplement our $R=0.4$ studies with
  a discussion of larger jet radii.}

\begin{figure}
  \centering
  \begin{minipage}[c]{0.48\linewidth}
  \includegraphics[width=\textwidth]{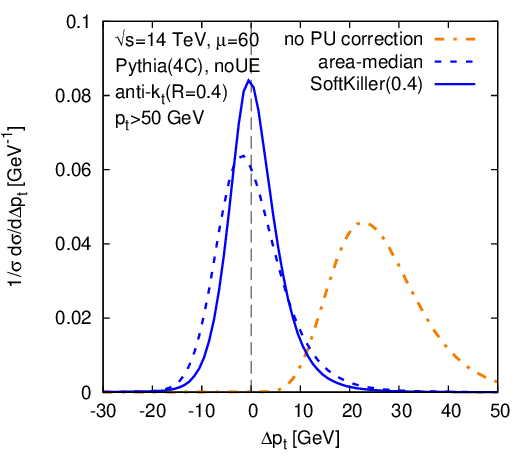}
  \end{minipage}
  \hfill
  \begin{minipage}[c]{0.48\linewidth}
  \caption{Performance of SoftKiller for 50 GeV jets and $\mu=60$
    Poisson-distributed pileup events. We plot the distribution of the
    shift $\Delta p_t$ between the jet $p_t$ after pileup removal and
    the jet $p_t$ in the hard event alone. Results are compared
    between the area-median approach and the SoftKiller. For
    comparison, the (orange) dash-dotted line corresponds to the situation
    where no pileup corrections are applied.}
  \label{fig:killer-pt-hdiff}
  \end{minipage}
\end{figure}

\begin{figure}
  \includegraphics[width=0.48\textwidth]{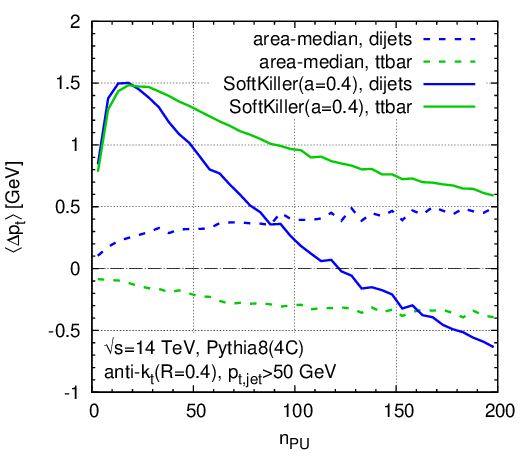}
  \hfill
  \includegraphics[width=0.48\textwidth]{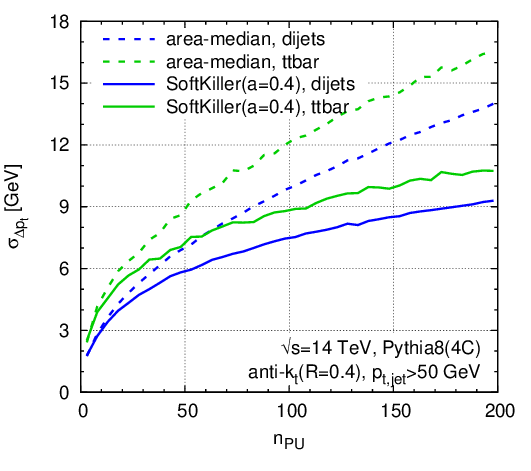}
  \caption{Performance of SoftKiller shown as a function of the pileup
    multiplicity and compared to the area-median subtraction method.
    Left: the average $p_t$ shift after subtraction, compared to the
    original jets in the hard event. 
    Right: the corresponding dispersion.}
  \label{fig:killer-v-npu}
\end{figure}
 
Next, let us compare the performance of the SoftKiller to that of
area--median subtraction.
Figure~\ref{fig:killer-pt-hdiff} shows the distribution of shift in
$p_t$, for (hard) jets with $p_t > 50\GeV$ in a dijet sample.
The average number of pileup events is $\mu = 60$, with a Poisson
distribution.
One sees that in the SoftKiller approach, the peak is about $30\%$
higher than what is obtained with the area--median approach and the
distribution correspondingly narrower.
The peak, in this specific case, is well centred on $\Delta p_t = 0$.

Figure~\ref{fig:killer-v-npu} shows the shift (left) and dispersion
(right) as a function of $n_\PU$ for two different samples: the $p_t >
50\GeV$ dijet sample (in blue), as used in
Fig.~\ref{fig:killer-pt-hdiff}, and a hadronic $t\bar t$ sample,
with a $50\GeV$ $p_t$ cut on jets (in green). 
Again, the figure compares the area--median (dashed) and SoftKiller
results (solid).
One immediately sees that the area--median approach gives a bias that
is more stable as a function of $n_\PU$.
Nevertheless, the bias in the SoftKiller approach remains between
about $-0.5$ and $1.5\GeV$, which is still reasonable when one
considers that, experimentally, some degree of recalibration is anyway
needed after area--median subtraction.
As concerns sample dependence of the shift, comparing $t\bar t$ v.\
dijet, the area--median and SoftKiller methods appear to have similar
systematic differences.
In the case of SoftKiller, there are two main causes for the sample
dependence: firstly the higher multiplicity of jets has a small effect
on the choice of $\ptcut$ and secondly the dijet sample is mostly composed
of gluon-induced jets, whereas the $t\bar t$ sample is mostly composed of
quark-induced jets (which have fewer soft particles and so lose less
momentum when imposing a particle $p_t$ threshold).
Turning to the right-hand plot, with the dispersions, one sees that
the the SoftKiller brings a significant improvement compared to
area--median subtraction for $n_\PU \gtrsim 20$.
The relative improvement is greatest at high pileup levels, where
there is a reduction in dispersion of $30-35\%$, beating the
$\sqrt{\nPU}$ scaling that is characteristic of the area--median
method. 
While the actual values of the dispersion depend a little on the
sample, the benefit of the SoftKiller approach is clearly visible for
both.

\begin{figure}
  \includegraphics[width=0.48\textwidth]{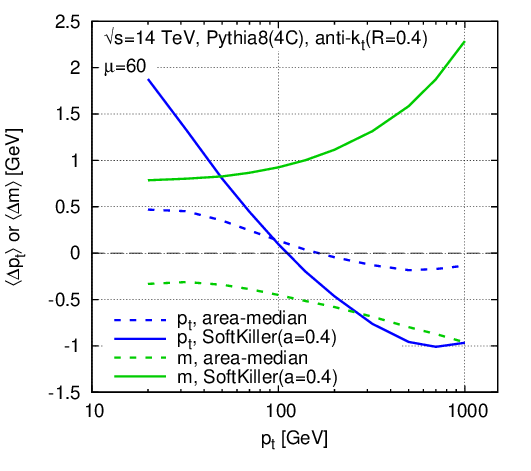}
  \hfill
  \includegraphics[width=0.48\textwidth]{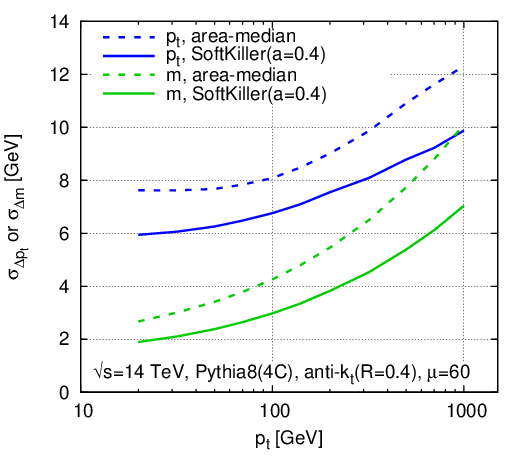}
  \caption{Performance of SoftKiller shown as a function of the hard
    jet $p_t$ (solid lines), with a comparison to the area-median
    subtraction method (dashed lines).
    Left: the average shifts for the jet $p_t$ (blue) and mass (green),
    after subtraction.  
    Right: the corresponding dispersions.  }
  \label{fig:killer-v-pt}
\end{figure}

Figure~\ref{fig:killer-v-pt} shows the shift (left) and dispersion
(right) for jet $p_t$'s and jet masses, now as a
function of the hard jet minimum $p_t$.
Again, dashed curves correspond to area--median subtraction, while solid
ones correspond to the SoftKiller results.
All curves correspond to an average of $60$ pileup interactions.
For the jet $p_t$ (blue curves) one sees that the area--median shift
ranges from $0.5$ to $0 \GeV$ as $p_t$ increases from $20\GeV$ to
$1\TeV$, while for SK the dependence is stronger, from about $2$ to
$-1\GeV$, but still reasonable.
For the jet mass (green curves), again the area--median
method\footnote{using a ``safe'' subtraction procedure
  that replaces negative-mass jets with zero-mass
  jets~\cite{Cacciari:2014jta}.} is more stable than 
SK, but overall the biases are under control, at the $1$ to $2\GeV$ level.
Considering the dispersions (right), one sees that SK gives
a systematic improvement, across the whole range of jet $p_t$'s. 
In relative terms, the improvement is somewhat larger for the jet mass
($\sim 30\%$) than for the jet $p_t$ ($\sim 20\%$).

\begin{figure}
  \includegraphics[width=0.48\textwidth]{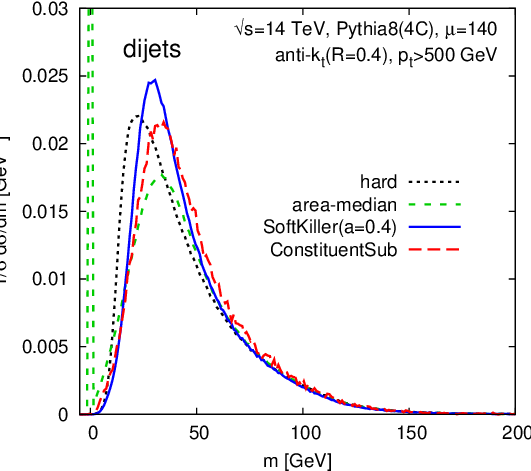}
  \hfill
  \includegraphics[width=0.48\textwidth]{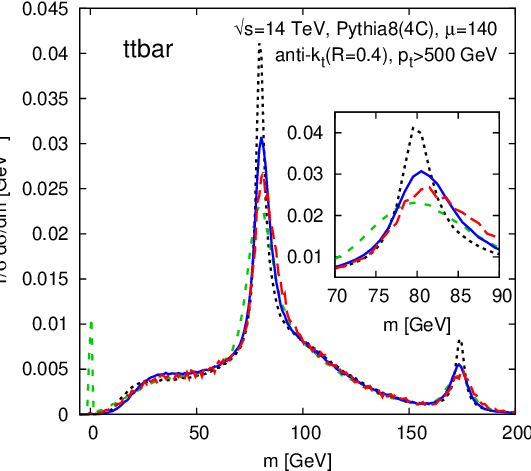}
  \caption{Performance of SoftKiller on jet mass reconstruction. We
    show the jet mass distribution after applying different pileup
    mitigation techniques. Left: dijet events, right:
    hadronically-decaying $t\bar t$ events.
}
  \label{fig:mass-spectra}
\end{figure}

Fig.~\ref{fig:mass-spectra} shows the actual mass spectra of $R=0.4$
jets, for two samples: a QCD dijet sample and a boosted $t\bar t$
sample.
For both samples, we only considered jets with $p_t > 500\GeV$ in the
hard event.
One sees that SK gives slightly improved mass peaks relative to the
area--median method and also avoids area--median's spurious peak at
$m=0$, which is due to events in which the squared jet mass came out
negative after four-vector area-subtraction and so was reset to zero.
The plot also shows results from the recently proposed Constituent
Subtractor method~\cite{Berta:2014eza}, using v.~1.0.0 of the
corresponding code from FastJet Contrib~\cite{FastJetContrib}.
It too performs better than area--median subtraction for the jet mass,
though the improvement is not quite as large as for
SK.\footnote{A further option is to use an ``intrajet killer'' that
  removes soft particles inside a given jet until a total $p_t$ of
  $\rho A_{\rm jet}$ has been subtracted. This shows performance
  similar to that of the Constituent Subtractor.}

\begin{figure}
  \includegraphics[width=0.48\textwidth]{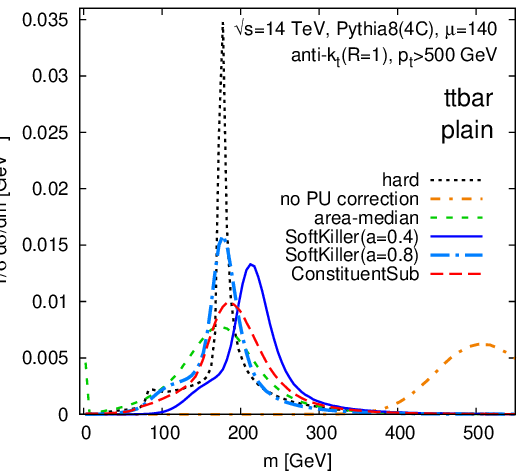}
  \hfill
  \includegraphics[width=0.48\textwidth]{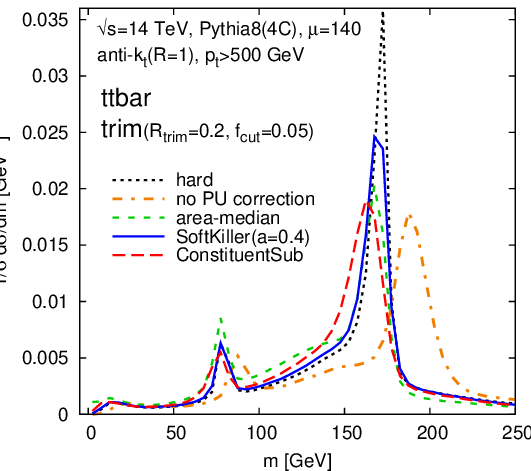}
  \caption{The results of the SoftKiller and other methods applied to
    large-$R$ jets. The left-hand plot is without grooming, the
    right-hand plot with grooming (trimming~\cite{trim} with $R_\text{trim}=0.2$
    and $f_\cut = 0.05$).
  }
  \label{fig:large-R}
\end{figure}

One might ask why we concentrated on $R=0.4$ jets here, given that
jet-mass studies often use large-$R$ jets.
The reason is that large-$R$ jets are nearly always used in
conjunction with some form of grooming, for example trimming,
pruning or filtering~\cite{trim,Ellis:2009su,filter}.
Grooming reduces the large-radius jet to a collection of small-radius
jets and so the large-radius groomed-jet mass is effectively a
combination of the $p_t$'s and masses of one or more small-radius
jets.

%
For the sake of completeness, let us briefly also study the SoftKiller
performance for large-$R$ jets.
Figure~\ref{fig:large-R} shows jet-mass results for the same $t\bar t$
sample as in Fig.~\ref{fig:mass-spectra} (right), now clustered with
the anti-$k_t$ algorithm with $R=1$.
The left-hand plot is without grooming: one sees that SK with our
default spacing of $a=0.4$ gives a jet mass that has better resolution
than area--median subtraction (or the ConstituentSubtractor), but a
noticeable shift, albeit one that is small compared to the
effect of uncorrected pileup.
That shift is associated with some residual contamination from pileup
particles: in an $R=0.4$ jet, there are typically a handful of
particles left from pileup, which compensate low-$p_t$ particles lost
from near the core of the jet.
If one substantially increases the jet radius without applying
grooming, then that balance is upset, with substantially more pileup
entering the jet, while there is only slight further loss of genuine
jet $p_t$.
To some extent this can be addressed by using the SoftKiller with a
larger grid spacing (cf.\ the $a=0.8$ result), which effectively
increases the particle $\ptcut$. 
This comes at the expense of performance on small-$R$ jets (cf.\
Fig.~\ref{fig:killer-scan}).
An interesting, open problem is to find a simple way to remove pileup
from an event such that, for a single configuration of the pileup
removal procedure, one simultaneously obtains good performance on
small-$R$ and large-$R$ jets.\footnote{As an example, the $p_t$
  threshold could be made to depend on a particle's distance from
  the nearest jet core; however this then requires additional
  parameters to define what is meant by a nearby jet core and to
  parametrise the distance-dependence of the $p_t$ cut.}

As we said above, however, large-$R$ jet masses are nearly always used
in conjunction with some form of grooming.
Fig.~\ref{fig:large-R} (right) shows that when used together with
trimming~\cite{trim}, SoftKiller with our default $a=0.4$ choice
performs well both in terms of resolution and shift.

\begin{figure}
  \includegraphics[width=\textwidth]{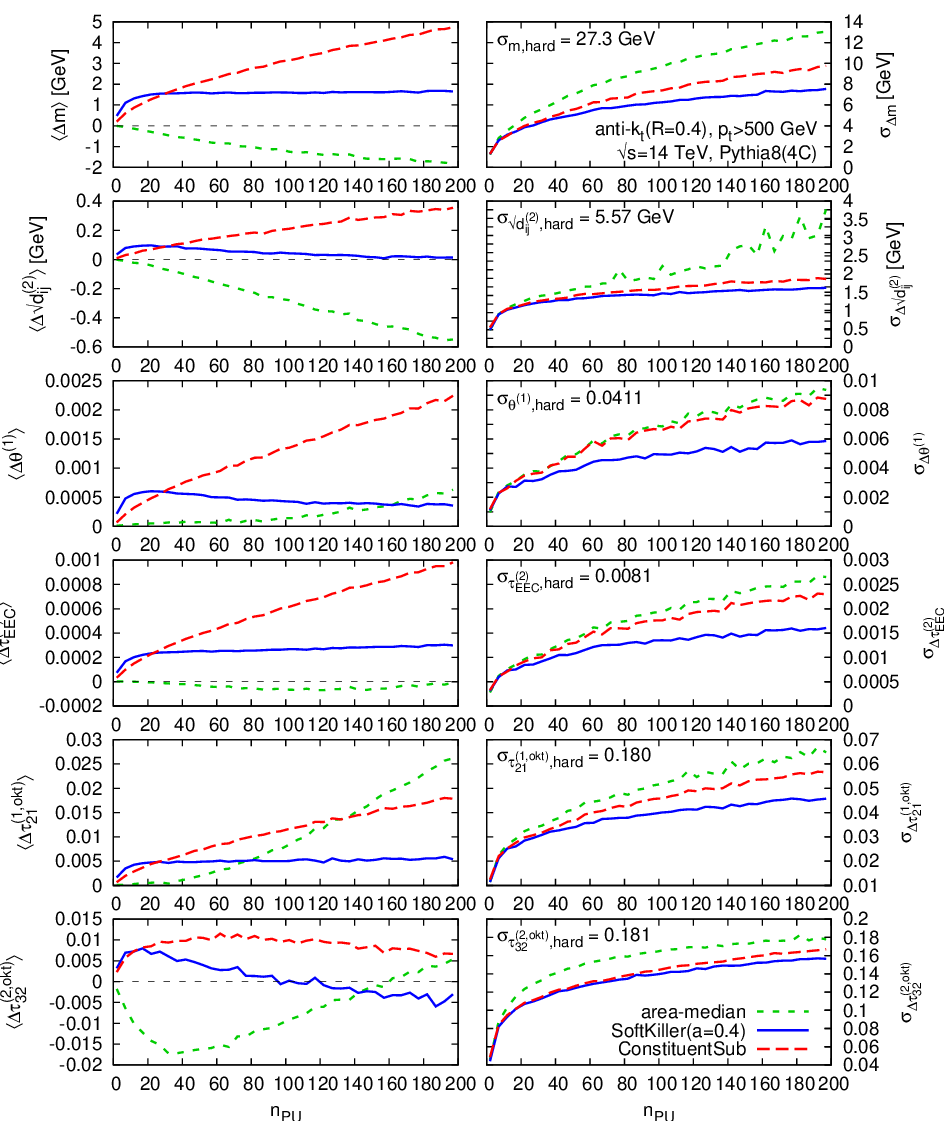}
  \caption{Performance of the SoftKiller on jet shapes, compared to
    area-median subtraction and the recently proposed Constituent Subtractor method
     \cite{Berta:2014eza}. All
    results are shown for dijet events with a 500 GeV $p_t$ cut on
    anti-$k_t$, $R=0.4$ jets.
    For comparison of the subtraction performance we also quote, for 
    each observable $X$,  $\sigma_{X,\rm hard}$, the dispersion of the
    distribution of the observable in the hard event.
    For $\tau_{21}$ there is the additional requirement (in the
    hard event) that the jet mass is above 30 GeV and for $\tau_{32}$
    we further impose that $\tau_{21}\ge 0.1$ (again in the hard
    event), so as to ensure infrared safety.
  \label{fig:shape-stats}
}
\end{figure}

Returning to $R=0.4$ jets, the final figure of this section,
Fig.~\ref{fig:shape-stats}, shows average shifts (left) and
dispersions (right) as a function of $n_\PU$ for several different jet
``shapes'': jet masses, $k_t$ clustering scales~\cite{kt}, the jet
width (or broadening or girth~\cite{Catani:1992jc,Berger:2003iw,Almeida:2008yp}), 
an energy-energy correlation
moment~\cite{Larkoski:2013eya} and the $\tau_{21}^{(\beta=1)}$ and 
$\tau_{32}^{(\beta=2)}$ N-subjettiness ratios~\cite{Thaler:2010tr}, 
using the exclusive $k_t$ axes with one-pass of minimisation.
Except in the case of the jet mass (which uses ``safe'' area
subtraction, as mentioned above), the area--median results have been
obtained using the shape subtraction technique~\cite{Soyez:2012hv}, as
implemented in v.~1.2.0 of the GenericSubtractor in FastJet Contrib.

As regards the shifts, the SK approach is sometimes the best, other
times second best.
Which method fares worst depends on the precise observable. 
In all cases, when considering the dispersions, it is the SK that
performs best, though the extent of the improvement relative to other
methods depends strongly on the particular observable.
Overall this figure gives us confidence that one can use the SoftKiller
approach for a range of properties of small-radius jets.

\section{Adaptation to CHS events and calorimetric events}
\label{sec:adapt-calor-towers}

It is important to verify that a new pileup mitigation method works
not just at particle level, but also at detector level. 
There are numerous subtleties in carrying out detector-level
simulation, from the difficulty of correctly treating the detector
response to low-$p_t$ particles, to the reproduction of actual detector
reconstruction methods and calibrations, and even the determination of which
observables to use as performance indicators.
Here we will consider two cases: idealised charged-hadron-subtraction,
which simply examines the effect of discarding charged pileup
particles; and simple calorimeter towers.

For events with particle flow~\cite{pflow} and charged-hadron
subtraction (CHS), we imagine a situation in which all
charged particles can be unambiguously associated either with the
leading vertex or with a pileup vertex.
We then apply the SK exclusively to the neutral particles, which we
assume to have been measured exactly.
This is almost certainly a crude approximation, however it helps to
illustrate some general features.

\begin{figure}
  \centering
  \includegraphics[width=0.48\textwidth]{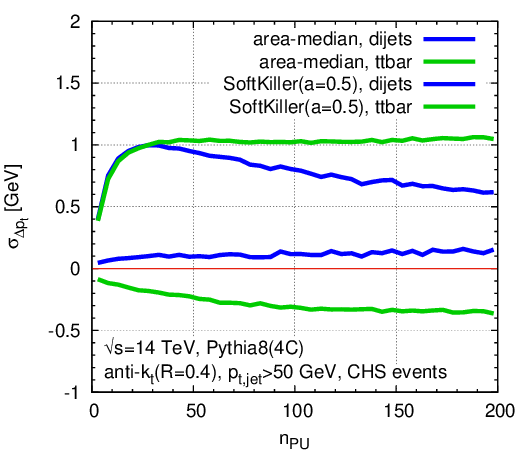}
  \hfill
  \includegraphics[width=0.48\textwidth]{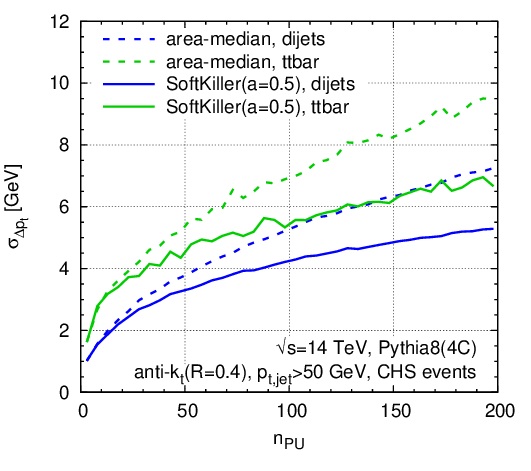}
  \caption{Same as Fig.~\ref{fig:killer-v-npu}, for events with
    charged-hadron subtraction (CHS). Note that the grid size used for
    the SoftKiller curves has been set to $a=0.5$.}
  \label{fig:chs-performance}
\end{figure}

One important change that arises from applying SK just to the neutral
particles is that there is a reduced contribution of low-$p_t$
hard-event particles. 
This means that for a given actual amount of pileup contamination (in
terms of visible transverse momentum per unit area), one can afford to
cut more aggressively, i.e.\ raise the $p_t^\cut$ as compared to the
full particle-level case, because for a given $p_t^\cut$ there will be a
reduced loss of hard-event particles.
This can be achieved through a moderate increase in the grid spacing,
to $a=0.5$.
Figure~\ref{fig:chs-performance} shows the results, with the shift
(left) and dispersion (right) for the jet $p_t$ in dijet and $t\bar t$
samples. 
The SK method continues to bring an improvement, though that
improvement is slightly more limited than in the
particle-level case. 
We attribute this reduced improvement to the fact that SK's greatest
impact is at very high pileup, and for a given
$n_\text{PU}$, SK with CHS is effectively operating at lower pileup
levels than without CHS.
A further study with our toy CHS simulation concerns lepton isolation
and is given in appendix~\ref{sec:isolation}.

Next let us turn to events where the particles enter calorimeter
towers.
Here we encounter the issue, discussed also in
appendix~\ref{sec:collinear-safety}, that SK is not collinear safe.
While we argue there that this is not a fundamental drawback from the point
of view of particle-level studies, there are issues at calorimeter
level: on one hand a single particle may be divided between two
calorimeter towers (we won't attempt to simulate this, as it is very
sensitive to detector details); on the other,
within a given tower (say $0.1\times0.1$) it is quite likely
that for high pileup the tower may receive contributions from multiple
particles.
In particular, if a tower receives contributions from a hard particle
with a substantial $p_t$ and additionally from pileup particles, the
tower will always be above threshold, and the pileup contribution will
never be removed.
There are also related effects due to the fact that two pileup
particles may enter the same tower.
To account for the fact that towers have finite area, we therefore
adapt the SK as follows.
In a first step we subtract each tower:
\begin{equation}
  \label{eq:tower-correction}
  p_t^\text{tower,sub} = \max\left(0,\; p_t^\text{tower} - \rho A^\text{tower}\right)\,,
\end{equation}
where $\rho$ is as determined on the event prior to any
correction.\footnote{We use our standard choices for determining
  $\rho$, namely the grid version of the area--median method,
  with a grid spacing of $0.55$ and rapidity scaling as discussed in
  Appendix~\ref{sec:rap-dep}. One could equally well use the same grid
  spacing for the $\rho$ determination as for the SoftKiller.}
This in itself eliminates a significant fraction of pileup, but there
remains a residual contribution from the roughly $50\%$ of towers
whose $p_t$ was larger than  $\rho A^\text{tower}$. 
We then apply the SoftKiller to the subtracted towers,
\begin{equation}
  \label{eq:ptcut-subtowers}
  p_t^{\cut,\text{sub}} = \underset{i \in \text{patches}}{\text{median}} \left\{ p_{ti}^{\text{tower,sub,}\max} \right\}\,,
\end{equation}
where $p_{ti}^{\text{tower,sub,}\max}$ is the $p_t$, after
subtraction, of the hardest tower in patch $i$, in analogy with
Eq.~(\ref{eq:ptmin}).
In the limit of infinite granularity, a limit similar to particle level,
$A^\text{tower}=0$. The step in Eq.~\eqref{eq:tower-correction} then has no
effect and one recovers the standard SoftKiller procedure applied to
particle level.

\begin{figure}
  \centering
  \includegraphics[width=0.48\textwidth]{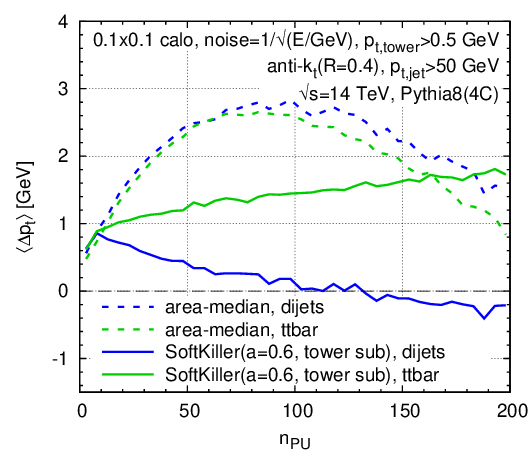}
  \hfill
  \includegraphics[width=0.48\textwidth]{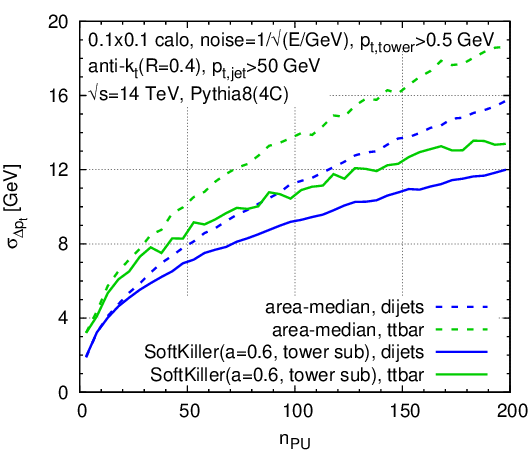}
  \caption{Same as Fig.~\ref{fig:killer-v-npu} for events with a
    simple calorimeter simulation. The SoftKiller
    was used here with a grid spacing of $a=0.6$ and includes the tower
    subtraction of Eq.~(\ref{eq:tower-correction}).}
  \label{fig:calo-performance}
\end{figure}

Results are shown in Fig.~\ref{fig:calo-performance}.
The energy $E$ in each $0.1\times 0.1$ tower is taken to have Gaussian
fluctuations with relative standard deviation $1/\sqrt{E/\!\GeV}$.
A $p_t$ threshold of $0.5\GeV$ is applied to each tower after fluctuations.
The SK grid spacing is set to $a = 0.6$. 
Interestingly, with a calorimeter, the area--median method starts to
have significant biases, of a couple of GeV, which can be attributed
to the calorimeter's non-linear response to soft energy.
The SK biases are similar in magnitude to those in
Fig.~\ref{fig:killer-v-npu} at particle level (note, however, the need
for a different choice of grid spacing $a$).
The presence of a calorimeter worsens the resolution both for
area-median subtraction and SK, however SK continues to perform
better, even if the improvement relative to area--median subtraction
is slightly smaller than for the particle-level results.

We have also investigated a direct application of the particle-level
SoftKiller approach to calorimeter towers, i.e.\ without the
subtraction in Eq.~(\ref{eq:tower-correction}). 
We find that the biases were larger but still under some degree of
control with an appropriate tuning of $a$, while the performance on
dispersion tends to be intermediate between that of area--median subtraction
and the version of SoftKiller with tower subtraction.

The above results are not intended to provide an exhaustive study of
detector effects. 
For example, particle flow and CHS are affected by detector
fluctuations, which we have ignored; purely calorimetric jet
measurements are affected by the fact that calorimeter towers are of
different sizes in different regions of the detector and furthermore
may be combined non-trivially through topoclustering.
Nevertheless, our results help illustrate that it is at least
plausible that the SoftKiller approach could be adapted to a full
detector environment while retaining much of its performance advantage
relative to the area--median method.

\section{Computing time}
\label{sec:computing-time}

The computation time for the SoftKiller procedure has two components:
the assignment of particles to patches, which is $\order{N}$, i.e.\
linear in the total number of particles $N$ and the determination of
the median, which is $\order{P \ln P}$ where $P$ is the number of
patches.
The subsequent clustering is performed with a reduced number of
particles, $M$, which, at high pileup is almost independent of the
number of pileup particles in the original event.
In this limit, the procedure is therefore expected to be dominated by
the time to assign particles to patches, which is linear in $N$.
This assignment is almost certainly amenable to being parallelised.

In studying the timing, we restrict our attention to particle-level
events for simplicity.
We believe that calorimeter-type extensions as described in
section~\ref{sec:adapt-calor-towers} can be coded in such a way as to
obtain similar (or perhaps even better) performance.

\begin{figure}[t]
  \centering
  \includegraphics[width=0.48\textwidth]{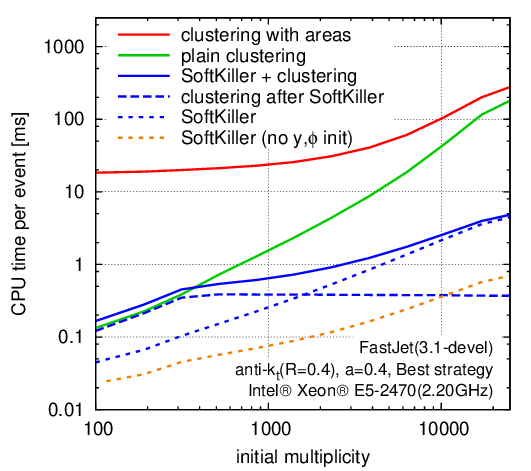}
  \hfill
  \includegraphics[width=0.48\textwidth]{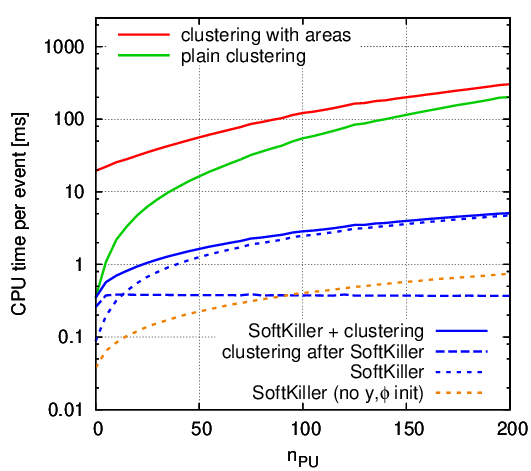}  
  \caption{Timings of the SoftKiller compared to standard clustering
    shown as a function of the number of particles in the event (left)
    or as a function of the number of pileup vertices (right). We
    compare the SoftKiller timings to the time to cluster the full
    event with (red) or without (green) jet area
    information. For the SoftKiller timings (in blue), we show
    individually the time spent to apply the SoftKiller to the event
    (dotted line), the time spent to cluster the resulting event
    (dashed line) and their sum (solid line).
    The orange dotted line corresponds to the SoftKiller timing when the
    particle's rapidity and azimuth have been precomputed.
  }
  \label{fig:timing}
\end{figure}

Timings are shown in Fig.~\ref{fig:timing} versus initial multiplicity
(left) and versus the number of pileup vertices (right).\footnote{
These timings have been obtained on an Intel Xeon processor, E5-2470
(2.20\,GHz), using a development version of FastJet 3.1, with the
``Best'' clustering strategy.
This has a speed that is similar to the public 3.0.6 version of
FastJet. Significant speed improvements at high multiplicity
  are planned for inclusion in the public release of FastJet 3.1,
  however they were not used here.} 
Each plot shows the time needed to cluster the full event and the time to
cluster the full event together with ghosts (as needed for area-based
subtraction).
It also shows the time to run the SoftKiller procedure, the time to
cluster the resulting event, and the total time for SK plus
clustering.

Overall, one sees nearly two orders of magnitude improvement in speed
from the SK procedure, with run times per event ranging from $0.2$ to $5\,$ms as
compared to $20$ to $300\,$ms for clustering with area information.
At low multiplicities, the time to run SK is small
compared to that needed for the subsequent clustering.
As the event multiplicity increases, SK has the effect of limiting the
event multiplicity to about $300$ particles, nearly
independently of the level of pileup and so the clustering time
saturates.
However the time to run SK grows and comes to dominate over the
clustering time.
Asymptotically, the total event processing time then grows linearly
with the level of pileup.
A significant part of that time (about $180\,\text{ns}$ per particle,
75\% of the run-time at high multiplicity) is taken by the
determination of the particles' rapidity and azimuth in order to
assign them to a grid cell. If the particles' rapidity and azimuth are
known before applying the SoftKiller to an event (as it would be the
case e.g. for calorimeter towers), the computing time to apply the
SoftKiller would be yet faster, as indicated by the dotted orange line
on Fig.~\ref{fig:timing}.

Because of its large speed improvement, the SoftKiller method has
significant potential for pileup removal at the trigger level. Since
SoftKiller returns an event with fewer particles, it will have a speed
performance edge also in situations where little or no time is spent
in jet-area calculations (either because Voronoi areas or fast
approximate implementations are used). This can be seen in
Fig.~\ref{fig:timing} by comparing the green and the solid blue
curves.

\section{Conclusions}
\label{sec:concl}

The SoftKiller method appears to bring significant improvements in
pileup mitigation performance, in particular as concerns the jet
energy resolution, whose degradation due to pileup is reduced by
$20-30\%$ relative to the area--median based methods.
As an example, the performance that is obtained with area--median
subtraction for 70 pileup events can be extended to 140 pileup events
when using SoftKiller.
This sometimes comes at the price of an increase in the biases on the
jet $p_t$, however these biases still remain under control.

Since the method acts directly on an event's particles, it
automatically provides a correction for jet masses and jet shapes, and
in all cases that we have studied brings a non-negligible improvement
in resolution relative to the shape subtraction method, and also (albeit
to a lesser extent) relative to the recently proposed Constituent
Subtractor approach.

The method is also extremely fast, bringing nearly two orders of
magnitude speed improvement over the area--median method for jet
$p_t$'s.
This can be advantageous both in time-critical applications, for
example at trigger level, and in the context of fast detector
simulations.

There remain a number of open questions.
It would be of interest to understand, more quantitatively, why such a
simple method works so well and what dictates the optimal choice of
the underlying grid spacing.
This might also bring insight into how to further improve the method. 
In particular, the method is known to have deficiencies when applied to
large-$R$ ungroomed jets, which would benefit from additional study.
Finally, we have illustrated that in simple detector simulations it is
possible to reproduce much of the performance improvement seen at
particle level, albeit at the price of a slight adaption of the method
to take into account the finite angular resolution of calorimeters.
These simple studies should merely be taken as indicative, and we
look forward to proper validation (and possible further adaptation)
taking into account full detector effects.

\textbf{Note added:} as this work was being completed, we became aware
of the development of another particle-level pileup removal method,
PUPPI~\cite{PUPPI}.
Initial particle-level comparisons at the 2014 Pileup
Workshop~\cite{PileupWorkshop} suggest that both bring comparable
improvements.

\section*{Acknowledgements}
We are grateful to Phil Harris, Peter Loch, David Miller, Filip
Moortgat, Ariel Schwartzman, and many others, for stimulating
conversations on pileup removal.
We are especially grateful to Filip Moortgat for his comments on the
manuscript. 
This work was supported by ERC advanced grant Higgs@LHC, by the French
Agence Nationale de la Recherche, under grant ANR-10-CEXC-009-01 and
by the EU ITN grant LHCPhenoNet, PITN-GA-2010-264564 
and by the ILP LABEX (ANR-10-LABX-63) supported by French state funds
managed by the ANR within the Investissements d'Avenir programme under
reference ANR-11-IDEX-0004-02.
GPS and GS wish to thank Princeton University and CERN, respectively,
for hospitality while this work was being carried out.

\appendix

\section{Collinear safety issues}
\label{sec:collinear-safety}

Collinear safety is normally essential in order to get reliable
results from perturbation theory.
One reaction to the SoftKiller proposal is that it is not
collinear safe, because it relies only on information about individual
particles' transverse momenta.
There are at least two perspectives on why this is not a severe issue.

The first relates to the intrinsic low-$p_t$ nature of the $p_t^\cut$,
which is typically of order $1-2\GeV$.
At these scales, non-perturbative dynamics effectively regulates the
collinear divergence.
Consider one element of the hadronisation process, namely resonance
decay, specifically $\rho \to \pi \pi$: if the $\rho$ has a $p_t$ of
order $2\GeV$, the rapidity-azimuth separation of the two pions is of
the order of $0.7-1$ (see e.g.\ Ref.~\cite{Cacciari:2014jta}).
Alternatively, consider the emission from a high-energy parton of a
gluon with a $p_t$ of the order of $1\GeV$: this gluon can only be
considered perturbative if its transverse momentum relative to the
emitter is at least of order a GeV, i.e.\ if it has an angle relative
to the emitted of order $1$.
Both these examples illustrate that the collinear divergence that is of
concern at parton level is smeared by non-perturbative
effects when considering low-$p_t$ particles.
Furthermore, the impact of these effects on the jet $p_t$ will remain of the
order of $\ptcut$, i.e.\ power-suppressed with respect to the scale of
hard physics.

The second perspective is from the strict point of view of
perturbative calculations.
One would not normally apply a pileup reduction mechanism in such a
context.
But it is conceivable that one might wish to \emph{define} the final
state such that it always includes a pileup and underlying event (UE)
removal procedure.\footnote{For example, so as to reduce prediction
  and reconstruction uncertainties related to the modelling of the UE
  (we are grateful to Leif L\"onnblad for discussions on this
  subject). This might, just, be feasible with area--median
  subtraction, with its small biases, but for the larger biases of SK
  does not seem phenomenologically compelling. Still, it is
  interesting to explore the principle of the question. }
Then one should understand the consequences of applying the
method at parton level.
Considering patches of size $0.5 \times \pi/6$ and particles with
$|y|<2.5$, there are a total of 120 patches; only when the
perturbative calculation
has at least $60$ particles, i.e.\ attains order $\alpha_s^{60}$, can
$p_t^\cut$ be non-zero; so the collinear safety issue would
enter at an inconceivably high order, and all practical fixed-order
parton-level calculations would give results that are unaffected by
the procedure.

Collinear safety, as well as being important from a
theoretical point of view, also has experimental relevance: for
example, depending on its exact position, a particle may shower
predominantly into one calorimeter tower or into two. Collinear
safety helps ensure that results are independent of these details. 
While we carried out basic detector simulations in
section~\ref{sec:adapt-calor-towers}, a complete study of the
impact of this type of effect would require full simulation and
actual experimental reconstruction methods (e.g.\ particle flow or
topoclustering).

\section{Rapidity dependence}
\label{sec:rap-dep}

\begin{figure}
  \centering
  \includegraphics[width=0.48\textwidth]{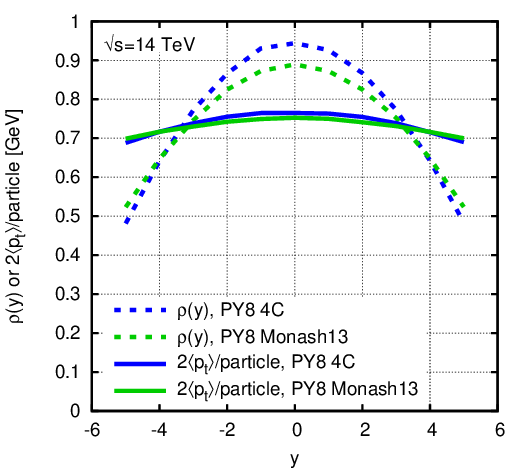}
  \hfill
  \includegraphics[width=0.48\textwidth]{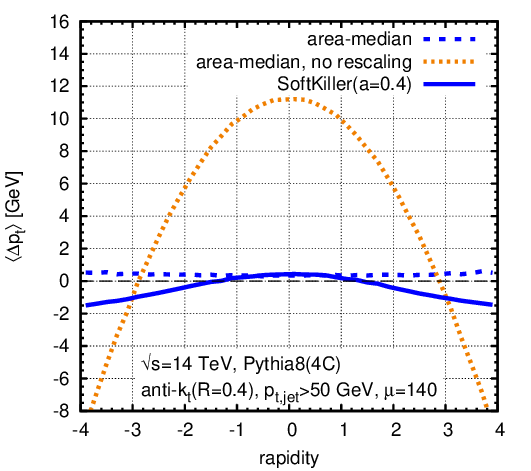}
  \caption{Left: rapidity dependence, in simulated zero-bias events,
    of the $p_t$ density per
    unit area ($\rho(y)$), and of the average $p_t$ per particle (scaled
    by a factor of $2$ for readability), comparing Pythia8's 4C and
    Monash~2013~\cite{Skands:2014pea} tunes.
    Right: the average shift in jet $p_t$ as a function of rapidity
    for the area--median method with rapidity rescaling (default in
    this paper), without rapidity rescaling, and for the SoftKiller as
    used throughout this paper, i.e.\ without rapidity rescaling.}
  \label{fig:rapidity-dep}
\end{figure}

One issue with the area--median method is that a global $\rho$
determination fails to account for the substantial rapidity dependence
of the pileup contamination.
Accordingly, the method is often extended by introducing an a-priori
determined function $f(y)$ that encodes the shape of the pileup's
dependence on rapidity $y$,
\begin{equation}
  \label{eq:rho-median-with-rap}
  \rho(y) = f(y) \, 
            \underset{i \in \text{patches}}{\text{median}} \left\{ \frac{p_{ti}}{A_i f(y_i)}\right\}\,.
\end{equation}
This is the approach that we have used throughout this
paper.\footnote{For Pythia8(4C) simulations, we use
  $f(y)=1.1685397-0.0246807 \, y^2+5.94119\cdot10^{-5}\, y^4$.}
The rapidity dependence of $\rho$, shown as the dashed lines in
Fig.~\ref{fig:rapidity-dep} (left), is substantial and therefore we
account for it through rescaling.
The figure shows two different tunes (4C and Monash~2013),
illustrating the fact that they have somewhat different rapidity
dependence.

The SoftKiller method acts not on the average energy flow, but instead
on the particle $p_t$'s. 
The solid lines in Fig.~\ref{fig:rapidity-dep} (left) show that the
average particle $p_t$ is nearly independent of rapidity.
This suggests that there may not be a need to explicitly account for
rapidity in the SK method, at least at particle level (detector
effects introduce further non-trivial rapidity dependence).

This is confirmed in the right-hand plot of
Fig.~\ref{fig:rapidity-dep}, which shows the rapidity-dependence of
the shift in the jet $p_t$ with the area--median and SK methods.
Our default area--median curve, which includes rapidity rescaling,
leads to a nearly rapidity-independent shift. 
However, without the rapidity rescaling, there are instead
rapidity-dependent shifts of up to $10\GeV$ at high pileup.
In contrast, the SK method, which in our implementation does not
involve any parametrisation of rapidity dependence, automatically
gives a jet $p_t$ shift that is fairly independent of rapidity, to
within about $2\GeV$.
We interpret this as a consequence of the fact (cf.\ the left-hand
plot of Fig.~\ref{fig:rapidity-dep}) that the average particle $p_t$
appears to be far less dependent on rapidity than the average $p_t$
flow.\footnote{In a similar spirit to
  Eq.~(\ref{eq:rho-median-with-rap}), one could also imagine
  introducing a rapidity-dependent rescaling of the particle $p_t$'s
  before applying SoftKiller, and then inverting the rescaling
  afterwards. Our initial tests of this approach suggest that it
  does largely correct for the residual SK rapidity
  dependence.}

\section{Monte Carlo tune dependence}
\label{sec:tune-dep}

\begin{figure}
  \includegraphics[width=0.48\textwidth]{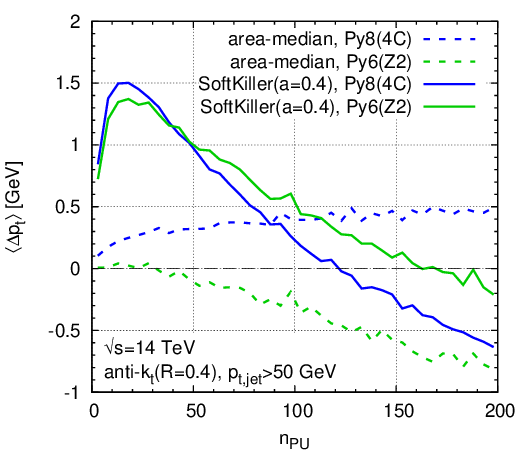}
  \hfill
  \includegraphics[width=0.48\textwidth]{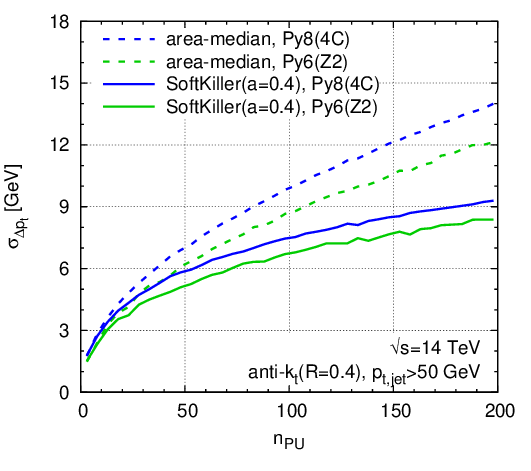}
  \caption{Performance of the area--median and SK pileup
    removal in the Pythia~6 Z2 tune as compared to the Pythia~8 4C
    tune that was our default throughout this paper.
    The results are the analogue of the dijet curves shown in
    Fig.~\ref{fig:killer-v-npu}. 
    Left: the average $p_t$ shift after subtraction, compared to the
    original jets in the hard event. 
    Right: the corresponding dispersion.}
  \label{fig:killer-v-npu-Z2-v-4C}
\end{figure}

While a full study of the dependence of the SK method on different Monte
Carlo tunes is beyond the scope of this article, we have briefly
verified that our conclusions are not affected by switching to another
widely used LHC tune, the Pythia~6~\cite{Sjostrand:2006za} Z2
tune~\cite{Field:2011iq}.
Fig.~\ref{fig:killer-v-npu-Z2-v-4C} compares the Pythia~6 Z2 results
for the jet $p_t$ offset and dispersion in a dijet sample with those from the
Pythia~8 4C tune that we used throughout the article.
While there are some differences between the two tunes, our main
conclusions appear unchanged. 
In particular, the average $p_t$ shift remains under control, and
there continues to be a significant improvement in the
resolution.\footnote{One may wonder about the stronger $n_\text{PU}$
  dependence for area--median subtraction with the Z2 tune as compared
  to 4C, however one should keep in mind that this corresponds to about just
  $5\,\text{MeV}$ per pileup vertex.}

\section{Impact of a fixed $\boldsymbol{p_t}$ cutoff}
\label{fixed-pt-cutoff}

\begin{figure}
  \centering
  \includegraphics[width=0.48\textwidth]{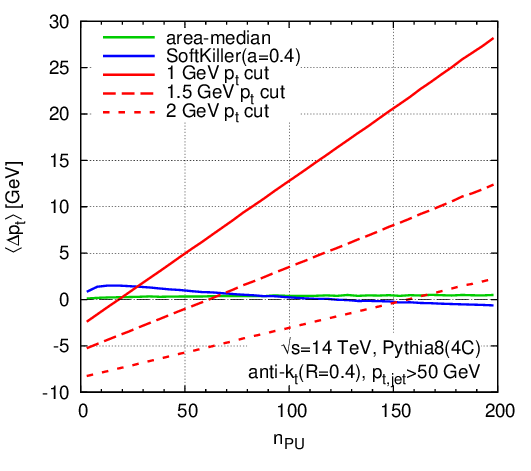}
  \hfill
  \includegraphics[width=0.48\textwidth]{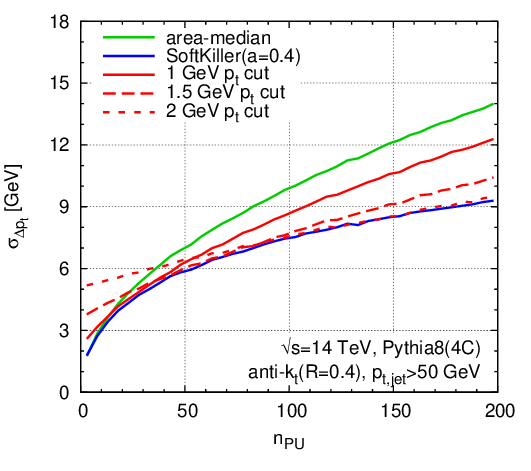}
  \caption{Comparison of area--median and SK results with those from a
    fixed $p_t$ cut.
    The average shift (left) and its dispersion (right) are shown as a
    function of $n_\text{PU}$ for dijet events, with a hard jet cut of
    $p_t > 50\GeV$.  }
  \label{fig:fixed-pt-cut}
\end{figure}

One key aspect of the SoftKiller approach is not simply that it
applies a $p_t$ cutoff, but rather that there is a straightforward
dynamical way of determining a $p_t$ cutoff, on an event-by-event
basis, that removes the bulk of the effects of pileup with modest
biases and improved dispersion.

For completeness, it is interesting to compare its performance to that
of a fixed $p_t$ cut.
Figure~\ref{fig:fixed-pt-cut} shows the shifts (left) and dispersions
(right), as a function of $n_\text{PU}$, as obtained for the
area--median method, the SoftKiller, and three fixed particle-level
$\ptcut$ values, $1\GeV$, $1.5\GeV$ and $2\GeV$.
For each of these fixed $\ptcut$ values, there is a value of $\nPU$
for which the shift in jet $p_t$ is zero, respectively $\nPU \simeq
20$, $60$ and $150$.
However, as soon as one moves away from that particular $\nPU$ value,
large biases appear.
Around the $\nPU$ that has zero bias for a given fixed $\ptcut$, the
dispersion of the $p_t$ shift is quite close to that obtained in the
SoftKiller approach. 
However, away from that $\nPU$ value, the dispersion becomes
somewhat worse.
Overall, therefore, the SoftKiller approach works noticeably better
than any fixed cut.

One further study that we have carried out is to parametrise the average
$\ptcut$ shown in Fig.~\ref{fig:killer-cut} (left) as a function of
$\nPU$, and to apply a $\ptcut$ that is chosen event-by-event
according to that event's actual value of $\nPU$.
We have found that this has performance that is similar to that of the
SoftKiller, i.e.\ SoftKiller's slight event-by-event adaptation of the
$\ptcut$ for a fixed $\nPU$ (represented by the 1-$\sigma$
dashed lines in Fig.~\ref{fig:killer-cut} (left)) does not appear to
be critical to its success.
This suggests that any approach that chooses an $\nPU$-dependent
$\ptcut$ so as to give a near-zero average $p_t$ shift may yield
performance on dispersions that is similar to that of SoftKiller.
From this point of view, SoftKiller provides an effective heuristic
for the dynamic determination of the $\ptcut$ value.

\section{Lepton isolation and (not) MET}
\label{sec:isolation}

Two non-jet-based quantities that suffer significantly from pileup
effects are lepton isolation and missing transverse energy (MET)
reconstruction.

Both potentially involve significant detector effects. 
For lepton isolation, we believe we may nevertheless be able to gain
some insight by considering a simplified scenario.
We consider isolation of hard leptons ($p_t > 25\GeV$) from $W$ decay
and also of hard leptons (with the same $p_t$ cut) from $B$-hadron
decays in events whose hard scattering was $gg$ or $q\bar q \to b\bar
b$.
The first sample provides genuinely isolated leptons, while the second
provides a sample of non-primary leptons, i.e.\ one important source
of lepton-production background that isolation is intended to
eliminate.
In both cases we use toy CHS events, as was described in
section~\ref{sec:adapt-calor-towers}. 

\begin{figure}
  \centering
  \includegraphics[width=0.48\textwidth]{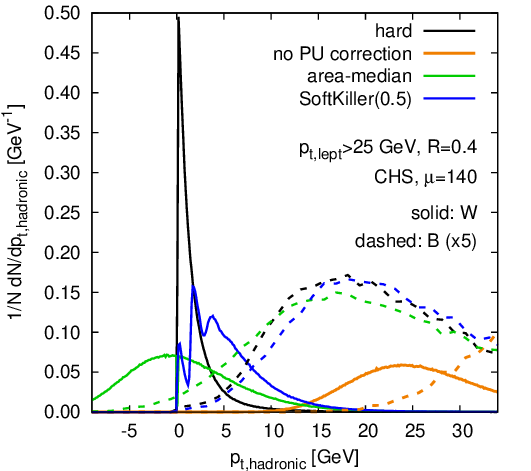}
  \hfill
  \includegraphics[width=0.48\textwidth]{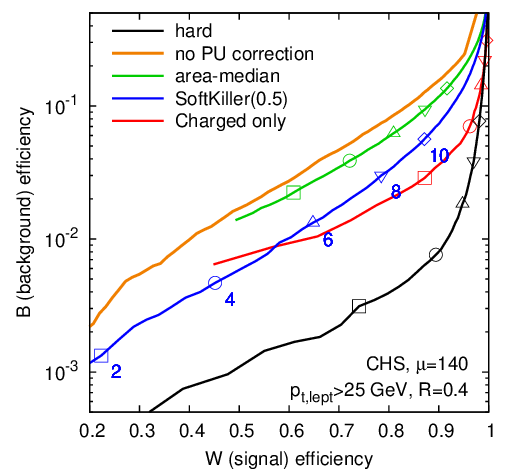}
  \caption{Left: simulated spectrum of (hadronic) $p_t$ in an $R=0.4$ cone
    around leptons with $p_t > 25\GeV$, for leptons from $W$ decay
    (solid curves) and from $B$-hadron decay (dashed curves); shown
    without pileup (black curves, ``hard'') and with $\mu=140$ pileup
    together with various pileup mitigation approaches.
    Right: resulting ROC curves showing the efficiency of isolating
    leptons from $B$-hadrons versus leptons from $W$'s.
    The different symbols indicate specific choices of isolation $p_t$
    cut (labelled in GeV for the blue SK curve).
  }
  \label{fig:isolation}
\end{figure}

Fig.~\ref{fig:isolation} (left) shows the $p_t$ contained in a cone of radius $0.4$ around
the lepton, with solid curves for leptons from $W$'s and dashed
curves for leptons from $B$ decays.
All curves except the black one (hard event only, i.e.\ no pileup)
correspond to events with a mean pileup of $\mu = 140$.
The orange curves illustrate how pileup severely shifts and smears the
distribution of $p_t$ around the lepton.
Area-median subtraction eliminates the shift, but gives only a
marginal improvement for the smearing.
SK gives somewhat more improvement as concerns the smearing, but has
the ``feature'' that there is a residual shift for the $W$ events, but
not for the $B$-hadron decays.
This difference is because $B$-hadron jets have some number of soft
particles that are removed by SK, compensating for the small residual
pileup left in by SK.
In contrast leptons from $W$'s tend to have few genuine soft particles
around them, so there is simply a net positive bias from the small
leftover PU. 
The peaks in the SK $W$-sample curve correspond to having $0$, $1$,
$2$, etc.\ residual pileup particles.

To establish how these characteristics translate to final performance,
one should examine the ROC curve for ``background'' efficiency
(i.e. for leptons from $B$'s) versus ``signal'' efficiency (i.e.\
leptons from $W$'s).
These are shown in the right-hand plot of Fig.~\ref{fig:isolation}, with the symbols
providing information about the isolation $p_t$ cut being used at a
given point on the curve.
Lower curves imply better performance.
One sees that uncorrected pileup (orange curve) significantly degrades
performance relative to the ``hard'' (i.e.\ no pileup) case. 
Area--median subtraction brings a small benefit and SK brings a
further moderate improvement. 
For a given isolation $p_t$ cut, the area--median approach gives a relatively
stable signal efficiency, while SK gives a more stable background
efficiency. 

A final comment about Fig.~\ref{fig:isolation} (right) concerns the red curve,
in which isolation is carried out just on the charged particles from
the primary vertex.
For signal efficiencies $\gtrsim 0.6$ this performs better than any
pileup-correction method (the exact value depends on the choice of $a$
for SK).
This highlights the point that it may be better to discard
pileup-contaminated information than it is to try to correct for the
large impact of pileup.\footnote{The very good performance of pure
  charged-particle isolation may be overoptimistic. 
  A shortcut in our simulation is that we assume that tracks from
  $B$-decays can be correctly associated with the primary vertex. 
  This may not be the case in a realistic environment.}
As well as discarding neutrals, one may also consider going to smaller
isolation radii, keeping in mind also recent theoretical progress in
understanding small-$R$ isolation and
jets~\cite{Catani:2013oma,Dasgupta:2014yra}.
The full optimisation over these various options should probably be
left to detailed experimental work.

Let us finally briefly comment on MET. 
With a perfect, infinite acceptance detector, pileup would have almost
no impact on MET, other than through the small fraction of neutrinos
present in pileup.
The large pileup-induced degradation in MET resolution that occurs in
real life is almost entirely a result of the interplay between the
detector (its acceptance and response) and pileup.
Without a detailed full detector simulation, we believe that it is
difficult for us to carry out a robust study of potential
improvements in MET reconstruction with SK-inspired methods.
Nevertheless, the fact that jet-area subtraction is used successfully
in ATLAS MET reconstruction~\cite{ATLASMET} suggests that the improvements
from SK may be of benefit also for MET.


\end{document}